\documentclass[10pt]{article}
\usepackage{graphicx}
\usepackage{amsmath}
\usepackage{subcaption}	
\usepackage{lineno}

\def\Title#1{\begin{center} {\Large #1 } \end{center}}
\def\Author#1{\begin{center}{ \sc #1} \end{center}}
\def\Address#1{\begin{center}{ \it #1} \end{center}}

\newcommand\pubblock{\rightline{\begin{tabular}{l} Proceedings of the Fifth Annual LHCP\\ \pubnumber\\
         \pubdate  \end{tabular}}}

\newenvironment{Abstract}{\begin{quotation} \begin{center} 
             \large ABSTRACT \end{center}\bigskip 
      \begin{center}\begin{large}}{\end{large}\end{center} \end{quotation}}

\newenvironment{Presented}{\begin{quotation} \begin{center} 
             PRESENTED AT\end{center}\bigskip 
      \begin{center}\begin{large}}{\end{large}\end{center} \end{quotation}}

\def\beq{\begin{equation}}
\def\eeq#1{\label{#1}\end{equation}}
\def\eeqn{\end{equation}}

\def\beqa{\begin{eqnarray}}
\def\eeqa#1{\label{#1}\end{eqnarray}}
\def\eeqan{\end{eqnarray}}

\let\bar=\overbar

\def\Dslash{\not{\hbox{\kern-4pt $D$}}}
\def\dslash{\not{\hbox{\kern-2pt $\del$}}}

\def\msb{{\bar{\ssstyle M \kern -1pt S}}}

\usepackage{color}
\usepackage{cite}

 \newcommand\pubnumber{ }

\newcommand\pubdate{\today}

\def\affiliation{
On behalf of the ALICE Experiment, \\
Istituto di Fisica Nucleare \\
Turin, Italy}

\begin{document}

\large
\begin{titlepage}
\pubblock

\vfill
\Title{Soft-QCD physics in pp and p-Pb with ALICE}
\vfill

\Author{ Valentina Zaccolo  }
\Address{\affiliation}
\vfill
\begin{Abstract}

At high collision energies as achieved at the LHC there are increasing contributions from hard processes, which can be computed with perturbative QCD. Nevertheless, particle production is still dominated by soft QCD with transferred momentum of a few GeV. These phenomena are described by non-perturbative phenomenology and challenge the theoretical models.
ALICE has measured several observables which target soft QCD, both in Run 1 and 2 proton-proton and proton-lead collisions at the LHC. A selection of results will be presented in these proceedings, focusing on the model comparisons and summarizing the understanding of soft QCD after almost 8 years of data taking.

\end{Abstract}
\vfill

\begin{Presented}
The Fifth Annual Conference\\
 on Large Hadron Collider Physics \\
Shanghai Jiao Tong University, Shanghai, China\\ 
May 15-20, 2017
\end{Presented}
\vfill
\end{titlepage}
\def\thefootnote{\fnsymbol{footnote}}
\setcounter{footnote}{0}
%

\normalsize 


\section{Introduction}

With higher collision energy at the LHC, the contribution from hard scatterings increases and more than one hard collision can occur, giving rise to Multiple Parton Interactions (MPI). 
Anyway, the majority of the processes observed are dominated by semi-hard and soft interactions: single, double and non-diffractive. The modelling of soft processes is challenging because of the non-perturbative aspects involved, making these measurements of high importance for tuning models. 
In the following, several observables will be presented and the common assessments discussed.
The ALICE detector is described elsewhere in detail \cite{Aamodt:2008zz}. For the event and centrality selection of the presented results, the V0 forward scintillators are used, as well as the Silicon Pixel Detector and the Alice Diffractive detector. For track reconstruction of charged particles, the Inner Tracking System (ITS) and the Time Projection Chamber (TPC) are used. The Forward Multiplicity Detector is used for counting particles at forward rapidity. The ITS is also used for vertexing, and the TPC for Particle Identification (PID). PID uses several other sub-detectors of ALICE: the Electromagnetic Calorimeter, the Transition Radiation Detector, the Time of Flight, and the Muon Spectrometer.

\section{Particle multiplicities}

\begin{figure}[t]
    \begin{subfigure}[h]{0.4\textwidth}
        \hspace{1cm}
        \includegraphics[width=\textwidth]{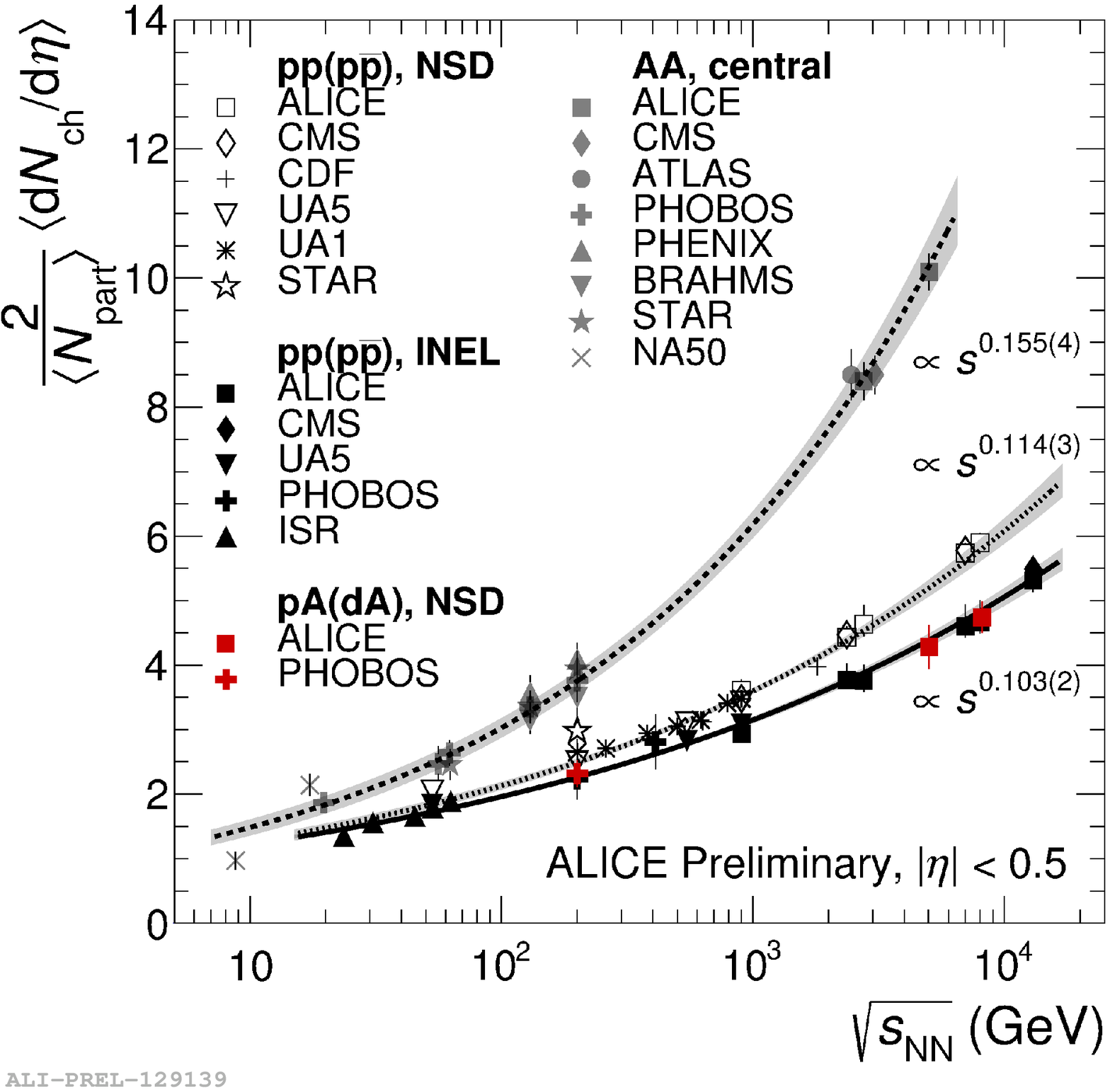}
    \end{subfigure}
  \begin{subfigure}[h]{0.4\textwidth}
       \hspace{1cm}
        \includegraphics[width=\textwidth]{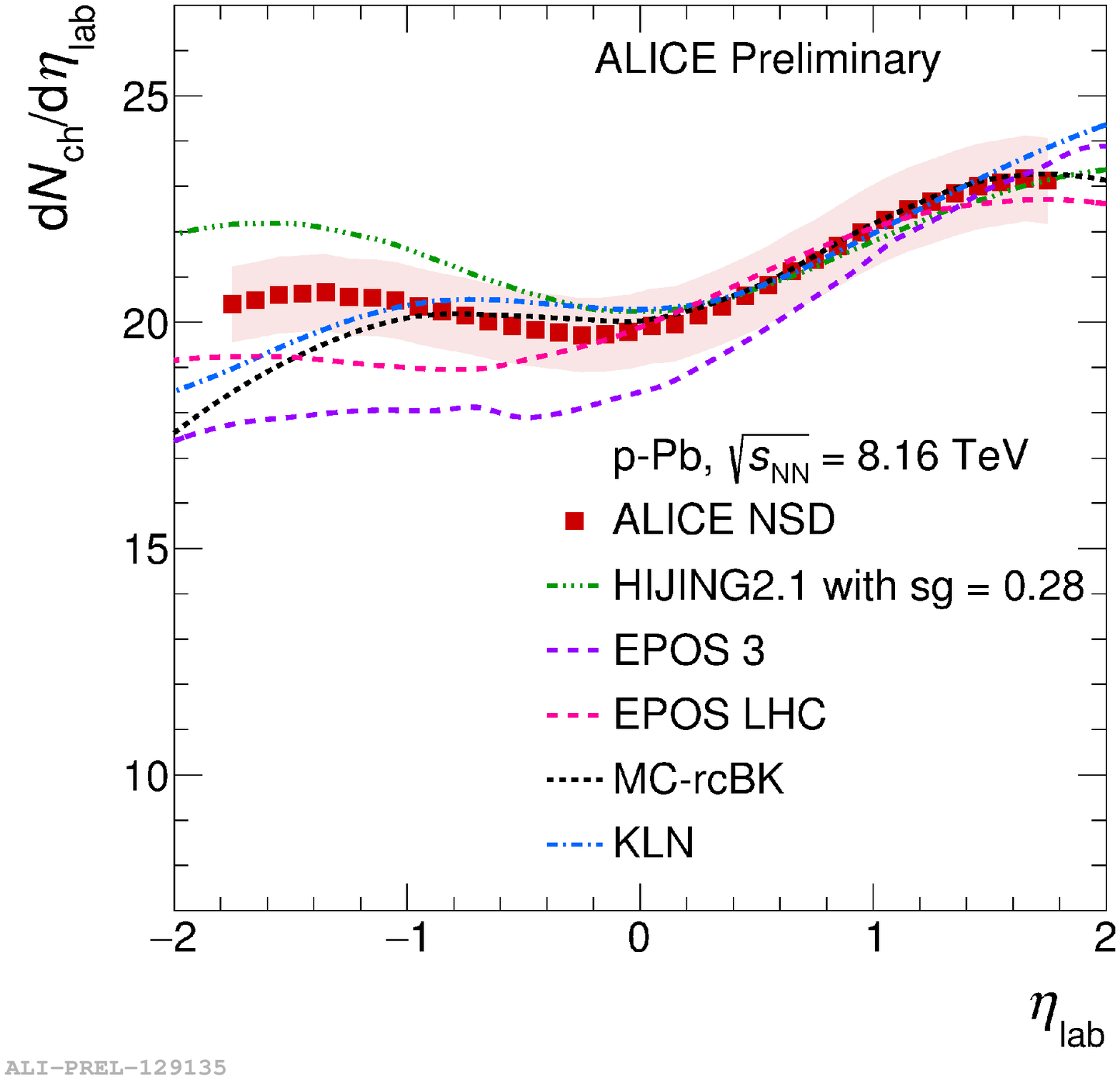}
	\end{subfigure}
\caption{Left: Charged-particle pseudorapidity density at midrapidity normalized to the number of \hbox{participants} as a function of $\sqrt{s_{\text{NN}}}$. Right: Pseudorapidity density of charged particles measured in non-single diffractive class p-Pb collisions at $\sqrt{s_{\text{NN}}}$=8.16 TeV, compared to theoretical predictions.}\label{fig:1}	
\end{figure}

Particle multiplicities are important for tuning theoretical models and as a reference for other measurements. 
In ALICE, we have measured both the pseudorapidity density $\text{d}N_{\text{ch}} / \text{d}\eta$  and the probability $\text{P}(N_{\text{ch}})$ as a function of the number of charged particles, in proton-proton, pp, collisions for Run 1 energies: $\sqrt{s}$ = 0.9 to 8 TeV, at central \cite{Aamodt:2009aa} and forward rapidities \cite{Zaccolo:2015udc}, and for Run 2 energy: 13 TeV \cite{Adam:2015pza}. 
Regarding proton-lead, p-Pb, collisions, new results for pseudorapidity density at $\sqrt{s_{\text{NN}}}$ = 8.16 TeV were presented and are shown in Fig.~\ref{fig:1}. 
On the left-hand side, the observable at midrapidity is scaled by half the average number of participants calculated with a Glauber model as a function of the energy in the center-of-mass system. The pA points agree with the pp inelastic (INEL) class, since the contribution from diffractive processes is negligible. It is also interesting to notice that the rise of AA points is much steeper with respect to pp and pA. 
In Fig.~\ref{fig:1} right, the distribution as a function of the pseudorapidity $\eta$ in the laboratory system is shown. The $\text{d}N_{\text{ch}} / \text{d}\eta_{\text{lab}}$ is asymmetric and the number of charged particles is higher in the Pb-going side, positive $\eta$. The distribution is compared to several models, which show a general good agreement in the Pb-fragmentation side \cite{Deng:2010mv, Werner:2013tya, Pierog:2013ria}. In the p side, models which assume gluon saturation,~MC-rkBK~\cite{Albacete:2012xq} and KLN~\cite{Dumitru:2011wq}, are favoured. 

\section{Underlying event}

\begin{figure}[t]
    \begin{subfigure}[h]{0.4\textwidth}
        \hspace{0.8cm}
        \includegraphics[width=\textwidth]{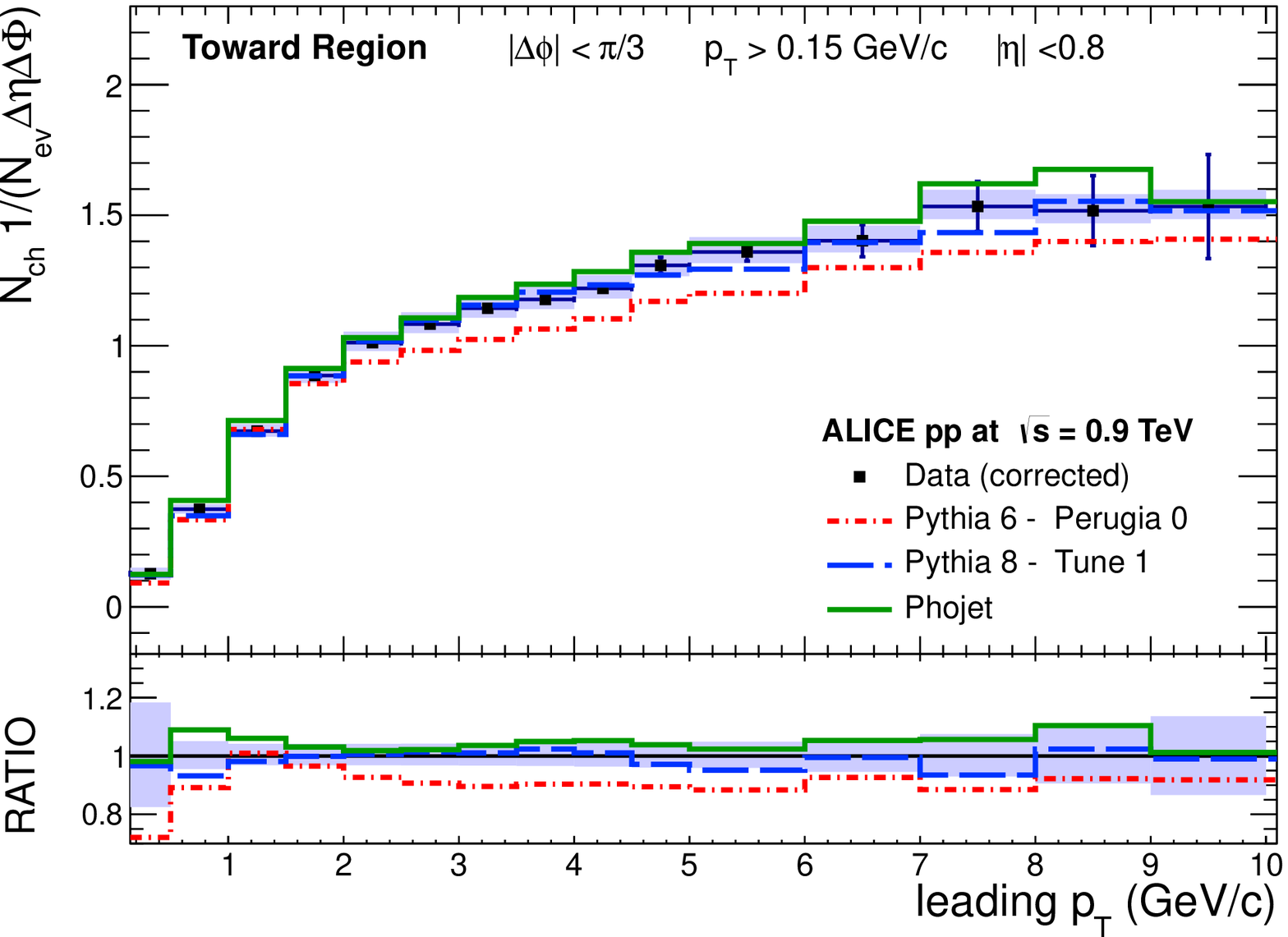}
    \end{subfigure}
  \begin{subfigure}[h]{0.4\textwidth}
       \hspace{1.3cm}
        \includegraphics[width=\textwidth]{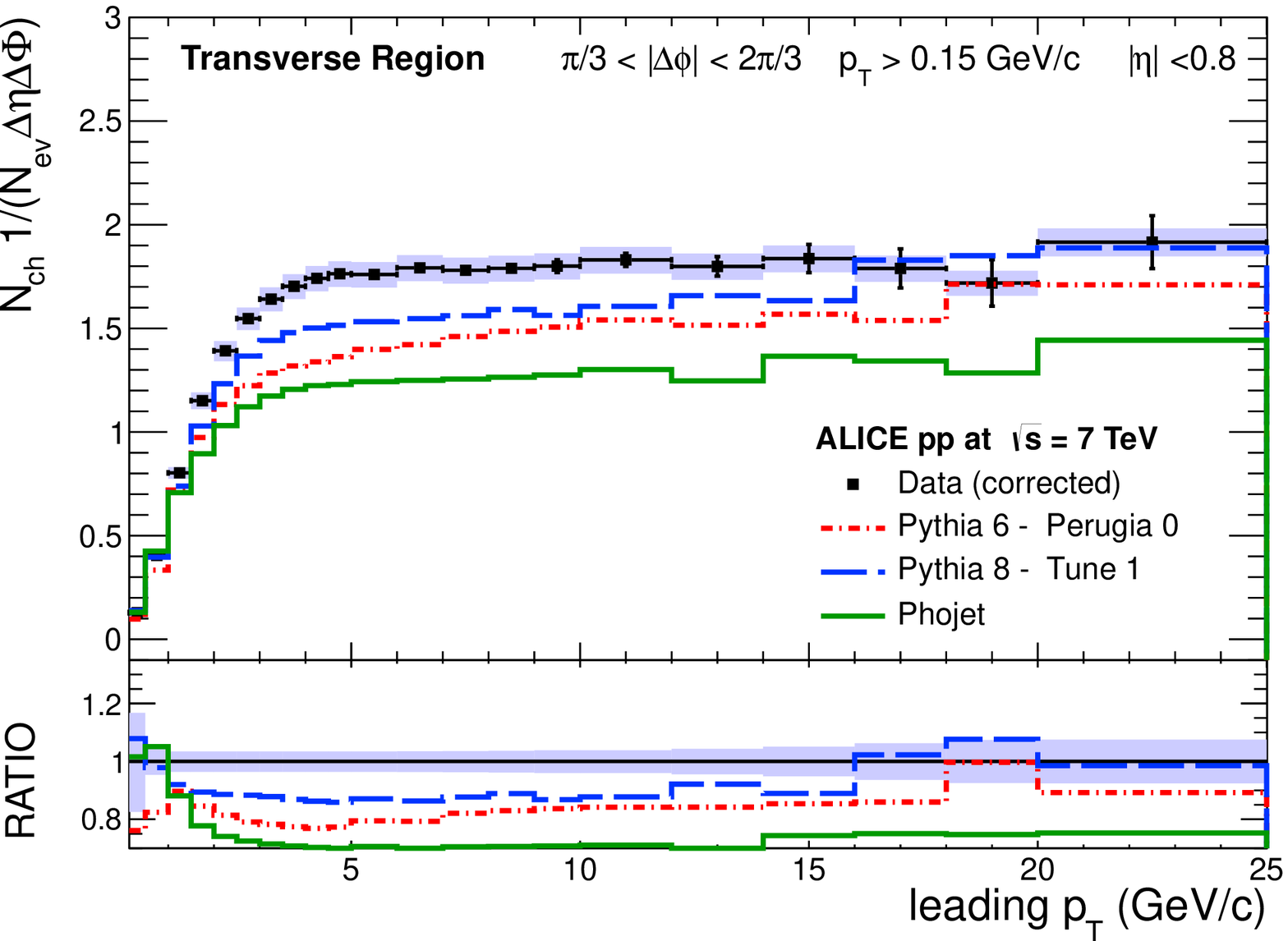}
	\end{subfigure}
\caption{Left: Number density in toward region at 7 TeV. Right: Number density in transverse region \cite{ALICE:2011ac}.}\label{fig:2}	
\end{figure}

A key measurement to separate soft and hard processes is the underlying event in different regions of the collision as a function of the leading-track momentum. 
In ALICE, we have measured it in pp collisions at $\sqrt{s}$ = 0.9 and 7 TeV \cite{ALICE:2011ac}.
The results for the average charged-particle density as a function of the $p_{\text{T}}$ of the leading track are shown in Fig.~\ref{fig:2} for the toward (left) and transverse (right) regions, for $p_{\text{T,min}}>$ 0.15~GeV.
The toward and away regions, with respect to the leading track, collect fragmentation products from hard scatterings and, there, the average particle density increases monotonically. 
The transverse-region measurement probes, instead, the underlying event and the particle density grows up to few GeV and than flattens. The plateau can be interpreted as a dependence on the MPI at low leading track $p_{\text{T}}$, while at higher leading track $p_{\text{T}}$ the hard processes do not influence the particle density any more.

\section{Multiple parton interactions}

\begin{figure}[t]
    \begin{subfigure}[h]{0.4\textwidth}
        \hspace{1cm}
        \includegraphics[width=\textwidth]{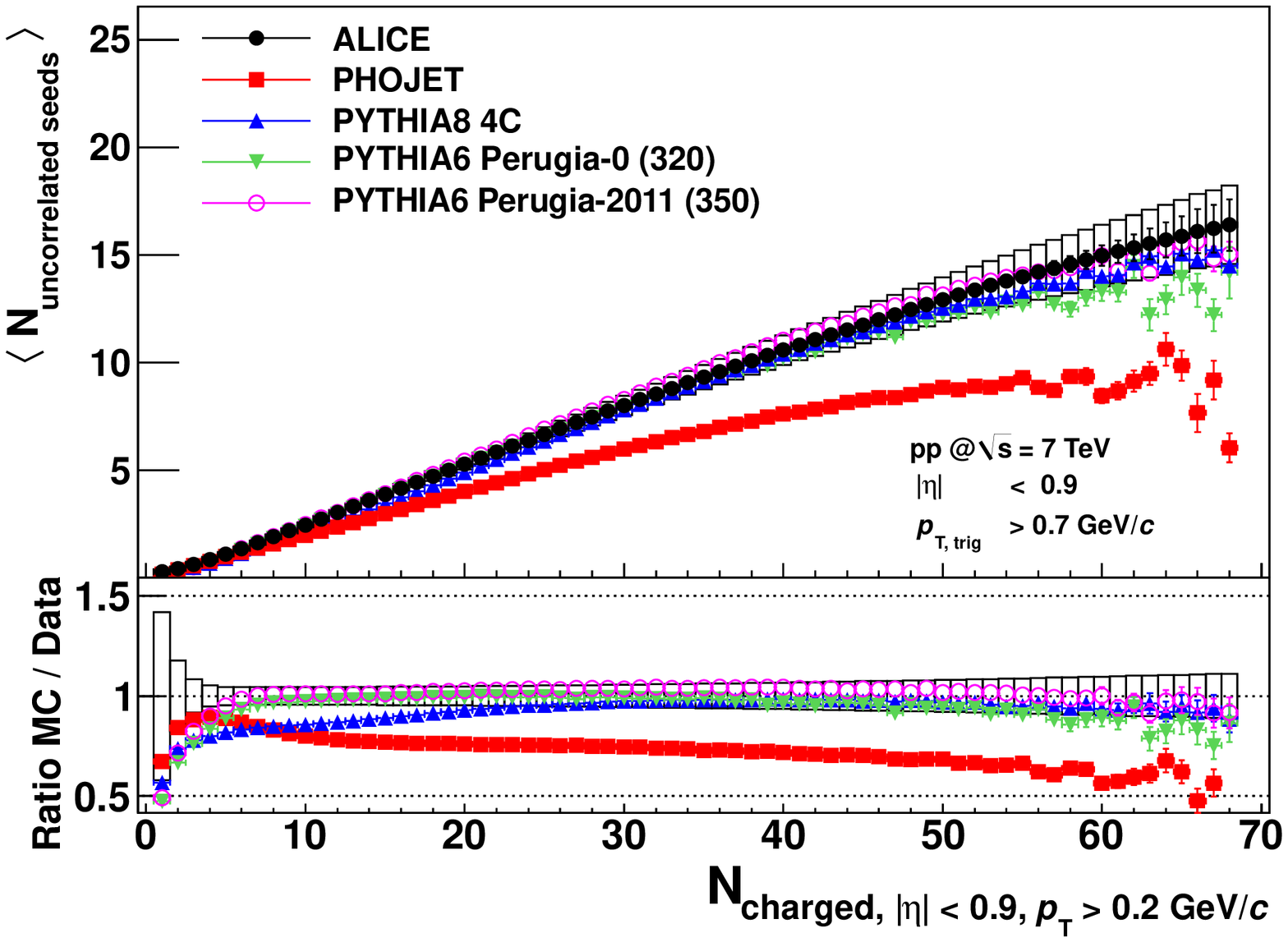}
    \end{subfigure}
  \begin{subfigure}[h]{0.4\textwidth}
       \hspace{1.3cm}
        \includegraphics[width=\textwidth]{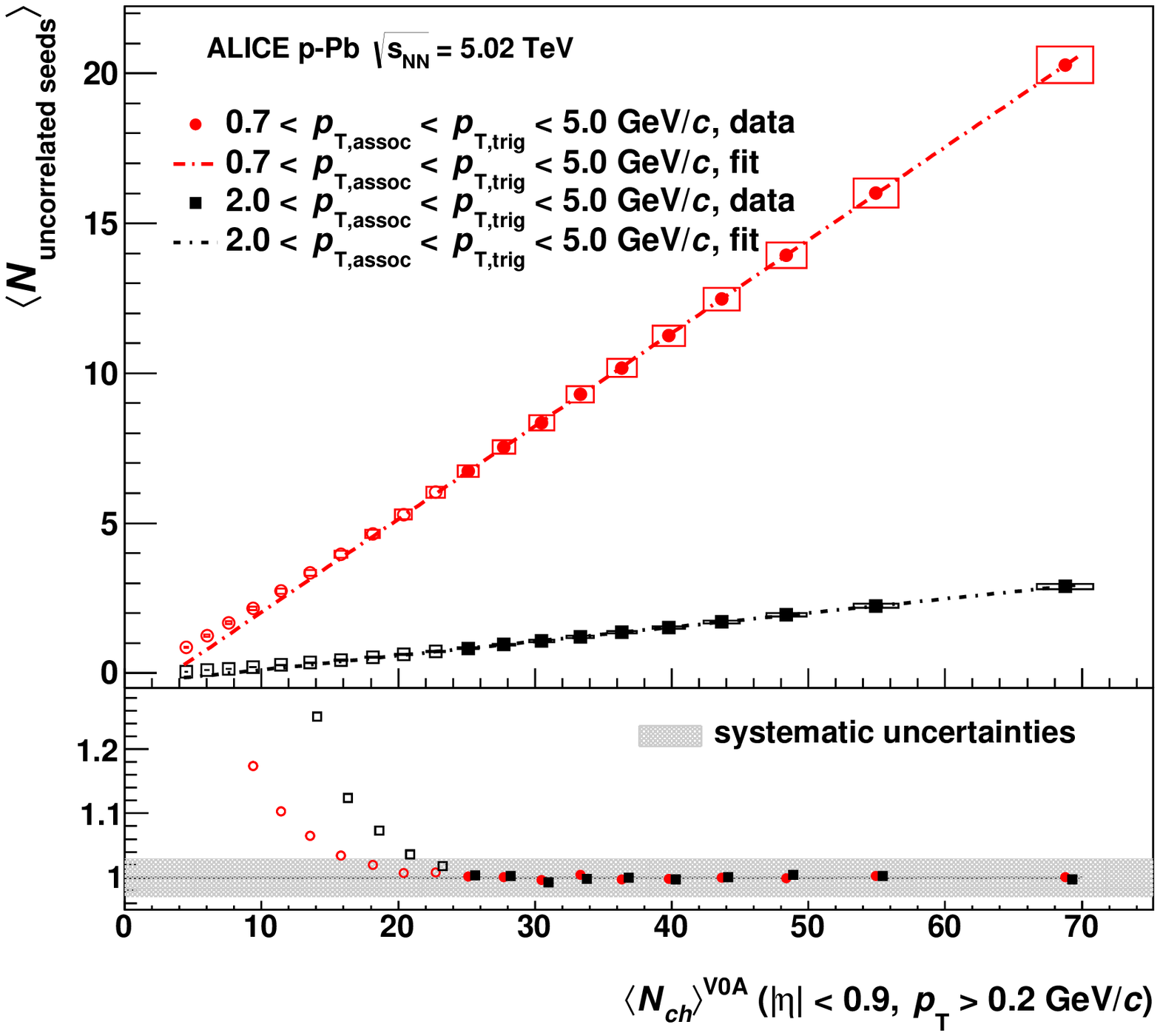}
	\end{subfigure}
\caption{Left: Average number of uncorrelated seeds measured in pp collisions at 7 TeV \cite{Abelev:2013sqa}. Right: Average number of uncorrelated seeds in p-Pb collisions in the 0-50\% event multiplicity class \cite{Abelev:2014mva}.}
\label{fig:3}	
\end{figure}

Wanting to study the number of MPI one can define the uncorrelated seeds.
In a parton-parton scattering two partons scatter back-to-back in the azimuthal angle. This creates two mini-jets, i.e.~jets at low $p_{\text{T}}$, with yield proportional to the number of MPI.
The MPI are roughtly proportional to the number of charged particles~\cite{Abelev:2013sqa}: $\langle N_{\text{uncorrelated seeds}}\rangle = \langle N_{\text{trig}}\rangle / \langle 1+N_{\text{associated, nearside}}+N_{\text{associated, awayside}}\rangle$. 
Particles are selected triggering using the first layer of the ITS and the V0 detector. $\langle N_{\text{trig}}\rangle$ depends on the number of semi-hard scatterings per event and the fragmentation properties of partons. 
Instead, the average number of uncorrelated seeds, $\langle N_{\text{uncorrelated seeds}}\rangle$, combines the average number of trigger particles with the nearside and awayside yield of trigger particles, reducing the fragmentation dependence and increasing the sensitivity to the number of scatterings per event. Starting from the measurement in pp at $\sqrt{s}$ = 7 TeV in Fig.~\ref{fig:3} left, one can notice that the $\langle N_{\text{uncorrelated seeds}}\rangle$ as a function of the charged particles increases up to a certain point where it saturates. This can be interpreted as a limit in the number of MPI. It is interesting to notice that the saturation behaviour is not present in p-Pb results at $\sqrt{s_{\text{NN}}}$=5.02 TeV, Fig.~\ref{fig:3} right, where the $\langle N_{\text{uncorrelated seeds}}\rangle$ continues to grow with $\langle N_{\text{ch}}\rangle$ \cite{Abelev:2014mva}. It will be important to check the behaviour at higher energies, although for pp collisions the saturation already starts at 0.9 TeV.

\section{Particle production}

\begin{figure}[t]
    \begin{subfigure}[h]{0.39\textwidth}
        \hspace{1cm}
        \includegraphics[width=\textwidth]{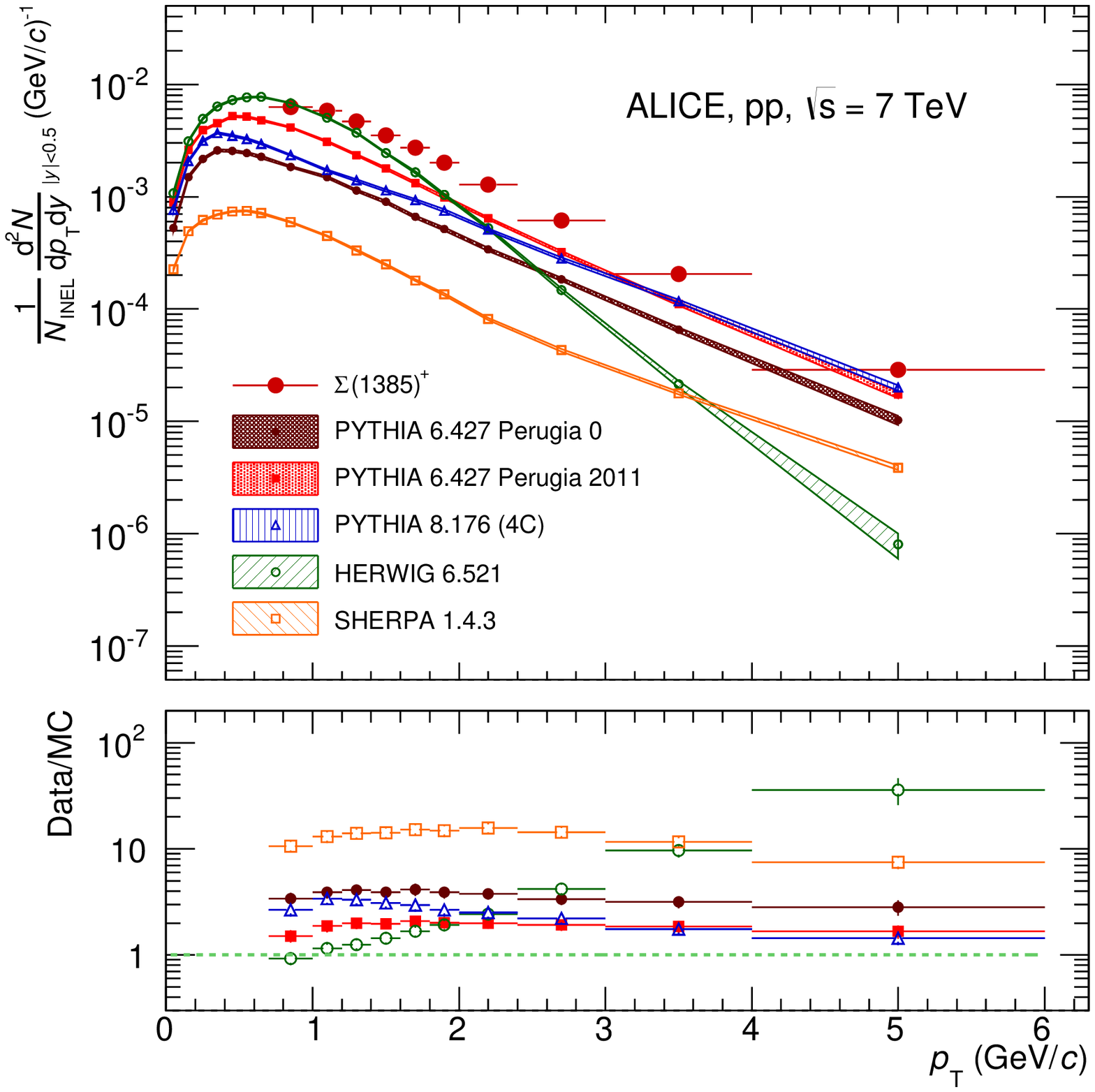}
    \end{subfigure}
  \begin{subfigure}[h]{0.41\textwidth}
       \hspace{1cm}
        \includegraphics[width=\textwidth]{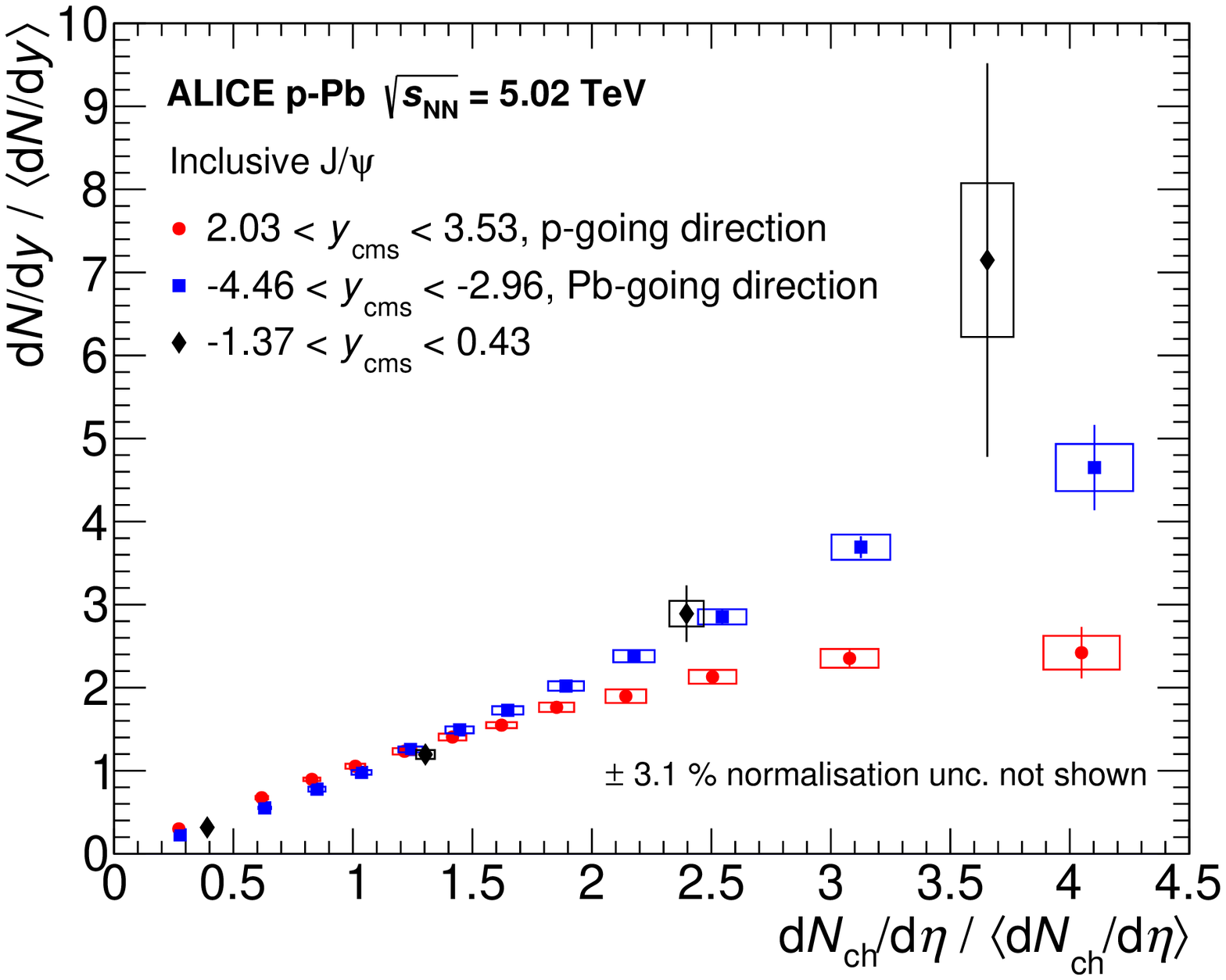}
	\end{subfigure}
\caption{left: Transverse momentum spectrum of $\Sigma(1385)^{+}$ in pp collisions at 7 TeV \cite{Abelev:2014qqa}. Right: Relative yield of inclusive $J/\psi$ mesons for three rapidity regions in p-Pb collisions at 5.02 TeV \cite{Adamova:2017uhu}.}
\label{fig:4}	
\end{figure}

As seen in the previous sections, the models do overall a good job in describing the observables, but one case where they generally fail is in the description of the strange-hadrons multiplicity dependence. This makes strangeness measurements crucial for tuning of Monte Carlo models.
Enhancement of strangeness has been observed by ALICE also in high-multiplicity pp collisions, challenging our understanding of small collision systems. 
One example is the recent measurement of the yields of strange and multistrange particles \cite{ALICE:2017jyt}. Indeed, this enhancement is one of the key observables to test the formation of Quark-Gluon Plasma in heavy-ion collisions. 
A representative strange-hadron measurement is the $\Sigma(1385)^{\pm}$ production in pp collisions, in Fig.~\ref{fig:4} left, where one notices that the Monte Carlo models compared to the results might differ also by an order of magnitude with respect to data \cite{Abelev:2014qqa}. Most recent versions of models, like PYTHIA 8 \cite{Sjostrand:2007gs}, agree better with the measurement. Other multiplicity-dependent results, which are relevant to study soft QCD, are the charmed and $J/\psi$ meson yields. 
In pp collisions, both at Run 1 and 2 energies, the D and $J/\psi$ yields as a function of the pseudorapidity density grow much faster than the diagonal \cite{Abelev:2012rz}. This can be interpreted as a result of multiplicity saturation, and, therefore, MPI saturation.
The same behaviour is observed in p-Pb collisions for the $J/\psi$ yield \cite{Adamova:2017uhu}. 
In Fig.~\ref{fig:4} right, one can see that in the forward rapidity region, which is also the region where the interaction is softer, there is a hint of saturation in the $J/\psi$ relative yield.

\section{Collectivity in small systems}

\begin{figure}[t]
\center
\includegraphics[width=0.4\textwidth]{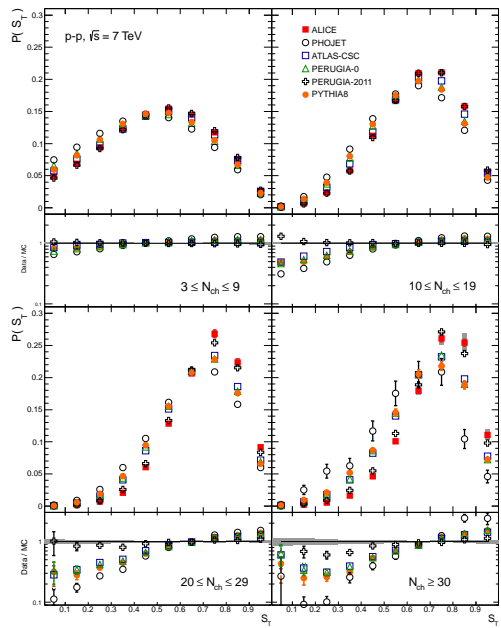}
\caption{Sphericity distributions in four bins of multiplicity in pp collisions at 7 TeV \cite{Abelev:2012sk}.}
\label{fig:5}	
\end{figure}

The event shape analysis is another good tool to check soft-QCD processes.
One of the observables measured in ALICE is the transverse sphericity, defined in terms of the eigenvalues of the transverse momentum matrix $S_{\text{T}}=2 \lambda_{2}/\lambda_{2}+\lambda_{1}$ \cite{Abelev:2012sk}. 
The  $S_{\text{T}}$ has a value that goes from 0, in the pencil-like events (hard processes), to 1, isotropic limit (soft processes).
An interesting measurement is the probability density, $\text{P}(S_{\text{T}})$, shown in Fig.~\ref{fig:5}. 
Looking at the results for $N_{\text{ch}}\geq30$ in the bottom-right pad, it can be concluded that the common Monte Carlo generators underestimate the production of isotropic events (higher $S_{\text{T}}$) but overestimate the production of jetty events. 
The measurement of another quantity, called spherocity $S_{0}=\pi^{2}/4 (\sum_{i}\overrightarrow{p_{\text{T}i}}\times\mathbf{\widehat{n}} / \sum_{i}p_{\text{T}i} )^2$, seems to be more effective in discriminating jets.

\section{Summary}

A wide set of soft-QCD results from ALICE have been shown.
In general, the charged-particle multiplicities and the underlying event are well described by models, at least on the level of 10-20\%. Given the complexity of non-perturbative soft-QCD description, this is a sound achievement. Still some additional work has to be done at forward rapidities where the collision is softer.
Several observables hint to saturation of Multiple Parton Interactions at high multiplicity and high $p_{\text{T}}$. Among these, we have presented a measurement of the underlying event, with the average charged-particle density in the toward region, the average uncorrelated seeds and the charmed and $J/\psi$ meson yields as a function of the multiplicity.
The models are particularly challenged when measuring strange hadrons, although progress has been made in recent tunes, which use LHC data.
Concluding, soft-QCD measurements at LHC have significantly improved our phenomenological understanding of high-energy collisions, but future measurements will allow us to further improve our understanding.

%
%
%
%
%
%
%
%
%
%
%


\begin{thebibliography}{99}


%
%

\bibitem{Aamodt:2008zz}
  K.~Aamodt {\it et al.} [ALICE Collaboration],
  JINST {\bf 3} (2008) S08002.
  doi:10.1088/1748-0221/3/08/S08002; 
  B.~Abelev {\it et al.} [ALICE Collaboration],
  Int.\ J.\ Mod.\ Phys.\ A {\bf 29} (2014) 1430044
  doi:10.1142/S0217751X14300440
  [arXiv:1402.4476 [nucl-ex]].
  
\bibitem{Aamodt:2009aa}
  K.~Aamodt {\it et al.} [ALICE Collaboration],
  Eur.\ Phys.\ J.\ C {\bf 65} (2010) 111
  doi:10.1140/epjc/s10052-009-1227-4
  [arXiv:0911.5430 [hep-ex]];
  K.~Aamodt {\it et al.} [ALICE Collaboration],
  Eur.\ Phys.\ J.\ C {\bf 68} (2010) 89
  doi:10.1140/epjc/s10052-010-1339-x
  [arXiv:1004.3034 [hep-ex]];
  K.~Aamodt {\it et al.} [ALICE Collaboration],
  Eur.\ Phys.\ J.\ C {\bf 68} (2010) 345
  doi:10.1140/epjc/s10052-010-1350-2
  [arXiv:1004.3514 [hep-ex]];
  J.~Adam {\it et al.} [ALICE Collaboration],
  Eur.\ Phys.\ J.\ C {\bf 77} (2017) no.1,  33
  doi:10.1140/epjc/s10052-016-4571-1
  [arXiv:1509.07541 [nucl-ex]].
  
\bibitem{Zaccolo:2015udc}
  V.~Zaccolo [ALICE Collaboration],
  Nucl.\ Phys.\ A {\bf 956} (2016) 529
  doi:10.1016/j.nuclphysa.2016.01.025
  [arXiv:1512.05273 [hep-ex]].
 
  
\bibitem{Adam:2015pza}
  J.~Adam {\it et al.} [ALICE Collaboration],
  Phys.\ Lett.\ B {\bf 753} (2016) 319
  doi:10.1016/j.physletb.2015.12.030
  [arXiv:1509.08734 [nucl-ex]].
  
\bibitem{Deng:2010mv}
  W.~T.~Deng, X.~N.~Wang and R.~Xu,
  Phys.\ Rev.\ C {\bf 83} (2011) 014915
  doi:10.1103/PhysRevC.83.014915
  [arXiv:1008.1841 [hep-ph]];
  R.~Xu, W.~T.~Deng and X.~N.~Wang,
  Phys.\ Rev.\ C {\bf 86} (2012) 051901
  doi:10.1103/PhysRevC.86.051901
  [arXiv:1204.1998 [nucl-th]].


\bibitem{Werner:2013tya}
  K.~Werner, B.~Guiot, I.~Karpenko and T.~Pierog,
  Phys.\ Rev.\ C {\bf 89} (2014) no.6,  064903
  doi:10.1103/PhysRevC.89.064903
  [arXiv:1312.1233 [nucl-th]];
  K.~Werner, I.~Karpenko, T.~Pierog, M.~Bleicher and K.~Mikhailov,
  Phys.\ Rev.\ C {\bf 82} (2010) 044904
  doi:10.1103/PhysRevC.82.044904
  [arXiv:1004.0805 [nucl-th]];
  H.~J.~Drescher, M.~Hladik, S.~Ostapchenko, T.~Pierog and K.~Werner,
  Phys.\ Rept.\  {\bf 350} (2001) 93
  doi:10.1016/S0370-1573(00)00122-8
  [hep-ph/0007198].
  
\bibitem{Pierog:2013ria}
  T.~Pierog, I.~Karpenko, J.~M.~Katzy, E.~Yatsenko and K.~Werner,
  Phys.\ Rev.\ C {\bf 92} (2015) no.3,  034906
  doi:10.1103/PhysRevC.92.034906
  [arXiv:1306.0121 [hep-ph]].
  
\bibitem{Albacete:2012xq}
  J.~L.~Albacete, A.~Dumitru, H.~Fujii and Y.~Nara,
  Nucl.\ Phys.\ A {\bf 897} (2013) 1
  doi:10.1016/j.nuclphysa.2012.09.012
  [arXiv:1209.2001 [hep-ph]];
  J.~L.~ALbacete and A.~Dumitru,
  arXiv:1011.5161 [hep-ph].
  
\bibitem{Dumitru:2011wq}
  A.~Dumitru, D.~E.~Kharzeev, E.~M.~Levin and Y.~Nara,
  Phys.\ Rev.\ C {\bf 85} (2012) 044920
  doi:10.1103/PhysRevC.85.044920
  [arXiv:1111.3031 [hep-ph]];
  D.~Kharzeev, E.~Levin and M.~Nardi,
  Nucl.\ Phys.\ A {\bf 730} (2004) 448
   Erratum: [Nucl.\ Phys.\ A {\bf 743} (2004) 329]
  doi:10.1016/j.nuclphysa.2004.06.022, 10.1016/j.nuclphysa.2003.08.031
  [hep-ph/0212316].
  
  %
\bibitem{ALICE:2011ac}
  B.~Abelev {\it et al.} [ALICE Collaboration],
  JHEP {\bf 1207} (2012) 116
  doi:10.1007/JHEP07(2012)116
  [arXiv:1112.2082 [hep-ex]].
  
\bibitem{Abelev:2013sqa}
  B.~Abelev {\it et al.} [ALICE Collaboration],
  JHEP {\bf 1309} (2013) 049
  doi:10.1007/JHEP09(2013)049
  [arXiv:1307.1249 [nucl-ex]].
  
\bibitem{Abelev:2014mva}
  B.~Abelev {\it et al.} [ALICE Collaboration],
  Phys.\ Lett.\ B {\bf 741} (2015) 38
  doi:10.1016/j.physletb.2014.11.028
  [arXiv:1406.5463 [nucl-ex]].
  
\bibitem{ALICE:2017jyt}
  J.~Adam {\it et al.} [ALICE Collaboration],
  Nature Phys.\  (2017)
  doi:10.1038/nphys4111
  [arXiv:1606.07424 [nucl-ex]].
  
\bibitem{Abelev:2014qqa}
  B.~Abelev {\it et al.} [ALICE Collaboration],
  Eur.\ Phys.\ J.\ C {\bf 75} (2015) no.1,  1
  doi:10.1140/epjc/s10052-014-3191-x
  [arXiv:1406.3206 [nucl-ex]].
  
\bibitem{Sjostrand:2007gs}
  T.~Sjostrand, S.~Mrenna and P.~Z.~Skands,
  Comput.\ Phys.\ Commun.\  {\bf 178} (2008) 852
  doi:10.1016/j.cpc.2008.01.036
  [arXiv:0710.3820 [hep-ph]].
  
\bibitem{Abelev:2012rz}
  B.~Abelev {\it et al.} [ALICE Collaboration],
  Phys.\ Lett.\ B {\bf 712} (2012) 165
  doi:10.1016/j.physletb.2012.04.052
  [arXiv:1202.2816 [hep-ex]];
  J.~Adam {\it et al.} [ALICE Collaboration],
  JHEP {\bf 1509} (2015) 148
  doi:10.1007/JHEP09(2015)148
  [arXiv:1505.00664 [nucl-ex]].
  
\bibitem{Adamova:2017uhu}
  D.~Adamova {\it et al.} [ALICE Collaboration],
  arXiv:1704.00274 [nucl-ex].
  
\bibitem{Abelev:2012sk}
  B.~Abelev {\it et al.} [ALICE Collaboration],
  Eur.\ Phys.\ J.\ C {\bf 72} (2012) 2124
  doi:10.1140/epjc/s10052-012-2124-9
  [arXiv:1205.3963 [hep-ex]].

  
  



\end{thebibliography}
\end{document}